# Structuring the hot advective accretion flow, as a result interaction of plasma with magnetic field

Krasimira Yankova[1], Lachezar Filipov[1]
*Space Research and Technology Institute - Bulgarian Academy of Sciences*
*Acad G. Bonchev Str., Bl. 1, 1113 Sofia, Bulgaria* **f7@space.bas.bg**

**Abstract**

*We present a magneto-hydrodynamic model developed for investigations of advective non-stationary, asymmetric Keplerian accretion disks in the normal magnetic field. The introduced model allows us to trace the evolution in different fixed moments and to get detailed description of the self-structuring of the disk.*

*We have introduced "meeting coefficients" that define the feedback. It determines the impact from nonlinear effects have over the structure of the flow.*

*We have obtained solutions for the radial structure of the disk for two crucial moments of its evolution.*

**Introduction**

The mass transfer in rotating systems with sufficiently large angular momentum leads to the formation of accretion disks, since the gas having angular momentum cannot fall directly onto the accreting-star. Then material will fall spirally, releasing some of its momentum together with the potential energy of a fixed orbit. The friction between neighbouring layers of the disk that have different densities is called viscosity (magnetic and kinetic viscosity). Viscosity is a mechanism that releases gravitational energy, transforming it into heat. Heat accelerates the movement of particles. Collisions between the particles lead to energy transfer from ions to electrons, thus turning it into radiation. This radiation occurs primarily in high energy ranges - X-rays and γ-rays.

The Standard Model: An accretion disk is a thermal machine. The disk is heated through viscosity that efficiently converts the mass of gas into energy. It is then cooled by radiation from both its surfaces. In other words, ions (the solid component) act as a heater and electrons are responsible for cooling.

Objects with high angular moment sufficient to form accretion disks are close binary systems [CBS] and active galactic nuclei [AGN]. Of all types of accretion, disk-like accretion is the most effective one. Up to 50% of the gas mass can be transformed into energy. Around the massive black holes, however, there are disks that cannot be cooled efficiently, because the release of energy in them is much faster than the speed of its transformation into radiation.

Cygnus X-1 is a close binary system and its accretor is a black hole. The effects that were registered were: high effective temperature, rapid variability in the different ranges, high luminosity, non-thermal spectrum, etc.; they cannot be explained by the standard model. Thermal radiation from the standard disc allows maximum temperature of $10^5$-$10^7$K, but the Cygnus X-1 indicates $T_e \sim 10^9$K in the interior.

Later optically thin additions (Chen et al. 1997) and two-temperature models (Subramanian et al. 1996) do not explain the leaps of temperatures registered at some objects either. A new idea was born. Advection can qualitatively modify the accretion theory. It fits in all existing models in such a way that preserves their main advantages.

Flows in advective disks are non-stationary (Chen et al. 1997) by nature and radial velocity can reach supersonic values. The time of inflow is short and dissipation works faster than diffusion. The disk cannot emit all the released energy. Some of it is converted into heat and remains inside the disk. Advection alters the balance in the disk (Beloborodov 1999, Bisnovatyi-Kogan 1998). It leads to a new kind of steady-state, but this steady-state is not EQUILIBRIUM. The value of the temperature can reach virial temperature (without free fall), but heavy protons remain non-relativistic. Advective accretion flows, in particular, are divided into two types (Beloborodov 1999, Bisnovatyi-Kogan 1998, Bisnovatyi-Kogan 1999, Chen et al. 1997, Narayan et al.1997, Narayan & Yi 1995): optically thick and optically thin flows. Advection in the optically thick disk takes place with higher accretion values. The density of flux increases dramatically as a result of which the opacity of the material increases as well. Diffusion slows down. Cooling is not effective. Advection smoothes out the thermal instability. Advective flow captures the radiation and brings it to the smaller radii along with the flow (Chen et al. 1997, Narayan & Yi 1995). In an optically thin disk high-temperature maintains ion-pressure high, and thus provides strong viscosity. Because of the low density, it is difficult to transmit the energy released to the $e^-$. Thus, the disk cannot be cooled effectively by radiation. The main energy remains in the accretion flow and increases $T_e$, which is in fact the cause of advection in these disks (Bisnovatyi-Kogan & Lovelace 2002).

The presence of a magnetic field contributes to efficient energy transfer from the ions to the cooler component. However, these disks are also difficult to cool because the field includes an additional mechanism of energy release. OHM dissipation increases the efficiency of radial accretion by 1.5 times. Dynamical time scale of $\sim \Omega^{-1}$, is shorter than the thermal time scale $\sim (\alpha + \alpha_m)\Omega^{-1}$ and the cooling is inefficient.

Most of the accretors show that they have a magnetic field. This is evident in the polarization of light and the annihilation line in the spectrum. For rotating stars with magnetic fields the electro-dynamic force is a stream-collimator (Kaburaki 1999). Even, if initially the flow is spherically symmetric, very soon it concentrates to the equatorial plane forming an accretion disk.

The first of these models (Hawley 2000, Hawley 2001, Hawley&Krolik 2001, Hawley&Krolik 2002) suggests that this disc is ideally conductive (the field does not penetrate the disk). A layer which penetrates into the magnetosphere, "freezes" in the field and moves with it. Then the "frozen stream" opposes the shifting, but the differential rotation causes a dynamo action, since there is no effective dissipation of the magnetic field in the disk.

In disks with finite conductivity, the material will not "freeze." Then diffusion, turbulent dissipation, the magnetic buoyancy and bipolar outflows provide for the weakening of the magnetic field.

Gas motion in accretion disks is sufficiently spiral-like, so that a Zeldovich dynamo necessarily occurs. In the case of differential rotation and an oscillating field, the dynamo will self-regulate by means of MRI which provides for the dynamo to remain in regime $D \approx D_{cr}$. This works in the cases of a small scale, when there is no stratification, and in the cases of a large scale when $v_a^2 \leq v_s^2$. 3D-simulations indicate that MRI could support the central field even in a global model (Brandenburg & Subramanian 2004). The difference is that in such a regime the main field does not alter significantly, only losses are compensated for.

Most of the authors of the MHD (Abramowicz et al.1996, Brandenburg &Campbell 1998, Brandenburg &Donner 1997, Campbell 1998, Fleming et al.2000, Gammie 2004; Matteo 1998) have investigated mainly dynamo-generated turbulence whereas other processes have been neglected. A common feature of most analytical models is that the authors choose to study one of the basic mechanisms with priority, ignoring others, as well as important effects such as the influence of the full pressure and own heating-cooling processes. Other authors (Armitage et al.1996, Bisnovatyi-Kogan 1998, Bisnovatyi-Kogan 1999, Bisnovatyi-Kogan & Lovelace 2002, Hawley & Balbus 2002) resort to methods of linear analysis and numerical simulation, because in this way the process of accretion can be illustrated better. However, the linearized problem can lead to the loss of significant nonlinear effects. Simulation codes sometimes cause non-physical effects that can affect the results so buffers have to be found.

That is why we propose a generalized model of a magnetized accretion disk with advection, which preserves the nonlinearity of the problem. In this paper, we present our studies of the interaction of the plasma flow with the magnetic field, as well as how this interaction affects the self-organization of the disk. The aim of these theoretical experiments is for us to follow the evolution and to reveal the mechanisms of the structuring of the disk-corona system and to interpret the high energy behavior of such sources.

Section 2 presents the construction of the model in Section 3 of our modifications.

Section 4 presents the approximate solution for the 2D-structure of the disc as well as the results and interpretation; Section 5 provides a summary of the conclusions.

**MODEL**

The basic equations of MHD of accretion - disk flow are: the continuity equation, equation of motion, equation of the magnetic induction, equation of heat balance and equation of state.

$$\frac{\partial \rho}{\partial t} + \nabla \cdot (\rho v) = 0 \tag{1}$$

$$\frac{\partial v}{\partial t} + v \cdot \nabla v = -\frac{1}{\rho}\nabla p - \nabla \Phi + (\frac{B}{4\pi\rho} \cdot \nabla)B + \vartheta \nabla^2 v \tag{2}$$

$$\frac{\partial B}{\partial t} = \nabla \times (v \times B) + \eta \nabla^2 B \tag{3}$$

$$\rho T \frac{\partial S}{\partial t} - \frac{\dot{M}}{2\pi r} T \frac{\partial S}{\partial r} = Q^+ - Q^- + Q_{mag} \tag{4}$$

$$p = p_r + p_g + p_m. \tag{5}$$

Where $\nabla \cdot v = 0$ and $\nabla \cdot B = 0$. (6)

The equations have been obtained in the conditions of two reference systems – coordinate system tied to the top of the flow and fixed coordinate system with centre in the accretor. Galilean transformations are applicable to the relative motion of both coordinate systems, one against the other. The flow will be considered non-relativistic because of $v^2/c^2 \sim 4.10^{-2} \ll 1$.

**Theoretical modelling**

We build a generalized model for the study of magneto-hydrodynamics of the advective accretion disk. For this purpose we construct non-stationary, non-axis-symmetric Keplerian disc in a normal magnetic field, based on the following assumptions:

- We have selected a black hole as a central object in our study. Its physics is most appropriate for surveying the magneto-rotational instabilities inside the disc (See interim results Iankova & Filipov 2005). Apart from that in his work (Khanna 1999) Khanna has shown that a weak field is always generated in the surroundings of a rotating black hole even in initially nonmagnetic plasma. We have selected the gravitational potential in the Pseudo-Newtonian form.

$\Phi = -\frac{GM}{\mathbf{r}-r_g}$ ; $r_g = \frac{2GM}{c^2}$; $\mathbf{r} = (r^2+z^2)^{1/2}$. It is a simple and convenient way to include the relativistic effects, which such a compact object exercises on the accretion in the consideration of a purely Newtonian flow (Abramowicz et al.1988).

- The velocity is introduced in the form $v = (v_r, r\Omega_k = r\sqrt{\frac{GM}{r(r-r_g)^2}}, 0)$, because the disk is Keplerian. Its initial value and direction are related to the degree of asymmetry of the disk (elliptical orbits) in the conditions of the axis-symmetric central potential; (it should be noted that this is not a specific object of study in this article, it is assumed that this parameter is known as it is considered an initial condition).

**MHD approximation**

Observations on skewed light near the event horizon of Cygnus X-1 show the presence of very strong magnetic fields. In polarized light new details about the warped space around the object are revealed. For seven years astronomers in Philippe Laurent`s team have been observing Cygnus X-1 through the IBIS telescope on board the Integral (satellite of the ESA). In 2012 they managed to observe the object at less than 800 km in diameter. Laurent who is a researcher at the Institute of Research into the Fundamental Laws of the Universe in Paris with the French Alternative Energies and Atomic Energy Commission's said: "Our results have shown for the first time that this unknown high-energy emission is strongly polarized, which implies that it should be produced by synchrotron radiation, a signature of a strong magnetic field at work close to the event horizon of the black hole" and added "There is no reason why other black hole binaries should not produce polarized light. We should observe this phenomenon in many other systems, also may be outside our galaxy."

It is sufficient to have a low residual dipole field which is then amplifies repeatedly by the gravito-magnetic dynamo. This deforms the Newton potential, as shown in (Kovar et al. 2010) But as the authors also show the pseudo-Newton potential gives surprisingly good results in the cases of BH with rotating dipole. This is the most appropriate approximation of the residual field, acquired mainly from the parent star. In this form it does not affect the space-time metric in the ergosphere, i.e. it does not affect the nature of the Newtonian flow.

Based on such observations and simulations we can assume that the main field $B_z$ of the accretor does not change with time. The main field is determined by the vertical component presented in the open form. This is due to the fact that in the cases of BH the magnetic field results from the initial distribution of the residual charge after the collapse and the subsequent activity of the gravito-magnetic dynamo. Therefore the field can be considered as a dipole oriented along the axis of rotation -- similar to the field of a star, but a lot weakly. Thus the MF remains sufficiently low so that we can consider the plasma in a weak magnetic field.

- In the equatorial plane $B_z = \frac{\mu}{r^3}$ – is independent of φ and t (Lipunov 1987). If the vertical field remains in the open form for all coordinates, this will not have a significant impact on the equations, because this will not introduce new parameters or add other differential terms.

- The fluid is neutral $\rho_e = 0$ – the concentration of free e⁻ in fully ionized hydrogen plasma is lower (<<) than the concentration that balances the charge;
- When the field is changing slowly, the shifting of the current $\frac{1}{c}\frac{\partial E_i}{\partial t}$ is negligible.
- With $\eta = \frac{c^2}{4\pi\sigma}$, we denote the electrical resistance.

Researchers believe that magnetic fields play an important role in the activities of black holes of all sizes: "It would not be a major surprise if all accretion disks rely on internal magnetic properties, at least partially."

The MF of an object can be private, central or external (galactic or from the star-donor). In the cases of own central field (as in this case) which is vertical to the plane of the disc, the angular momentum in the disk increases inward toward the centre. This effect can be compensated for by powerful jets, which take away the moment (Kuncic et al 2004). But until it reaches the central funnel the redistribution of the moment has an impact on the development of the flow. Accelerating rotation releases higher quantities of energy and thus has a direct impact (by entropy and advection and through entropy on advection) on the restructuring of the disc, in its capacity of an open system. In the case of optically thick flows that correspond to sub-Eddington accretion, the radiation is trapped inside the helical flow and therefore the stream is radiatively inefficient. (See, for example Igumenshchev &Abramovich 2000).

**Advection hypothesis - advective term**

Most authors consider advective-dominated sub-and super-Eddington flows. In our case, in our disc, advection is not the dominant mechanism. More popular models suggest flow deformation: 1) in the form of a sharp increase in the radial velocity and a significant decrease of orbital velocity to sub-Keplerian values. This is rotation of the vector of velocity. The physical meaning of the vector's rotation in models with radial advection is that the action of $\partial v_i/\partial t$ is ignored. ; or 2) orbital advection (Fabian et al.2012) for low- magnetic discs, where the maximum speed of sound is dominated by the Keplerian rotation. An orbital advection is a superposition of an orbital speed vector that is added to the Keplerian velocity. At this type of advection, the nature of the process requires the linearization in the model equations.

In our research, in contrast to these models, we provide the advection in the form of the complete advective term (7), which is naturally produced in the equations describing the flow dynamics:

$$\frac{\partial(\rho v_i)}{\partial t} + \frac{\partial}{\partial x_j}(\rho v_i v_j) = \rho\left(\frac{\partial v_i}{\partial t} + v_j \frac{\partial v_i}{\partial x_j}\right) = \rho \frac{Dv_i}{Dt} \tag{7}$$

Operator $\dfrac{D}{Dt}$ defines the advective term. In the equation of motion $\dfrac{\partial(\rho v_i)}{\partial t} + \dfrac{\partial}{\partial x_j}(\rho v_i v_j) = -\dfrac{\partial P_{ij}}{\partial x_j} + \dfrac{\partial M_{ij}}{\partial x_j} - \rho\dfrac{\partial \Phi}{\partial x_i}$, the left side is transformed $v_i\left(\dfrac{\partial \rho}{\partial t} + \dfrac{\partial(\rho v_j)}{\partial x_j}\right) + \rho\dfrac{\partial v_i}{\partial t} + \rho v_j\dfrac{\partial v_i}{\partial x_j}$ and then: $\dfrac{\partial(\rho v_i)}{\partial t} + \dfrac{\partial}{\partial x_j}(\rho v_i v_j) = \rho\left(\dfrac{\partial v_i}{\partial t} + v_j\dfrac{\partial v_i}{\partial x_j}\right) = \rho\dfrac{Dv_i}{Dt}$ is obtained.

This is not an increase of radial velocity. This is in fact shifting of the average flow as a flux with velocity $v_i$ in any direction. There is no rotation or lengthening of the vector, although this can be considered as particular cases (and it works well in the other models).

The more general interpretation, however, is that the decision as a whole is transferred to smaller radii. This is physically more probable in the case of such a powerful attraction by black holes. The orbital rate of course changes, but this is due to the new orbit and it is only indirectly connected with advection. Yet again the orbital velocity is Keplerian, because in the case of such a packet transfer, the flow does not change its nature. With a normal disk field the dipole term in the equation of motion $B_r B_\varphi$ creates the conditions for radial advection (Campbell et al. 1998) – i.e. this determines the direction of displacement of the middle course: In this context the importance of orbital scale is close in terms of its significance to the radial scale. Radial pulling into a black hole can reach a speed which is about half the rate of free fall, without weakening the orbital rotation; in fact the exact opposite is true, it is accelerated due to the redistribution of the momentum. Then, viscose and dynamic time-scale (which are comparable) is shorter than the thermal scale and cooling is inefficient.

Advection in similar conditions can work for relatively lower temperatures in the outer regions of the disk. Early appearances guarantee that the flow will remain optically thick at temperatures of order (2-s) which is higher than the normally accepted one.

- The disk is geometrically thin, optically thick, with no self-gravitation i.e. the fluid is incompressible;
- One-temperature and totally ionized plasma. The temperature in the flow is of the same order as the ion temperatures since the presence of the field gurantees the effective transport of energy towards the cooling component.

In the equation of heat balance, advection draws in the heat inside the disc and provides energy for structural-formation at the small radii.

The radial gradient of entropy is the advective term in heat balance. It can be presented in the form $-\dfrac{\dot{M}}{2\pi r}T\dfrac{\partial S}{\partial r} = -2H\rho v(-\dfrac{3}{2}\dfrac{B^2}{8\pi\rho r})$, because the advection is a direct result of the interaction of the magnetic field with plasma.

Cooling is in fact connected with the vertical gradient of entropy in a 2D-disk and has the form $Q^- = \dfrac{acT^4}{3\tau}$.

The sign of the entropy determines the basic criterion for development, equilibrium and stability of the disc. If it does not cool efficiently with time, the energy which is kept in disk in the form of heat, decreases gradient of entropy towards the centre. Negative entropy gradients create conditions for the absorption of energy from existing instabilities and thus stimulate feedbacks in the rearranging of the disc. In this way the internal structure of the disk is changing gradually in order to regulate the dynamic quasi-steady state of the disk. Such a state is relatively stable, but too far from equilibrium.

**Viscosity**

In this case we select the same type of viscose dissipation as in the standard model: $Q^+ = \vartheta \rho H \left( r \dfrac{\partial \Omega}{\partial r} \right)^2$.

The dynamo generation of turbulence is investigated in the following publications (Brandenburg & Campbell 1998, Brandenburg &Donner 1997, Campbell et al.1998, Kaburaki1999). The difference between the speeds of rotation of the plasma in the disk $\Omega$ and the speed of rotation of stars $\Omega_s$ and/or the shift, in the cases of different objects, generates an azimuthal component $B_\varphi$ (or the disturbance $b_\varphi$) of the magnetic field (Brandenburg &Campbell 1998). Movement in the disk (accretion, deccretion) stretches the magnetic field lines and generates $B_r$ and rotation twists them in the form of magnetic loops. Reconnection is presented by the formation of two opposing close loops in the neutral configuration. For smaller $l$ (local vortices scale) the current density $j = (c/4\pi)\nabla \times B \sim cB_r/4\pi l$ has a very high value and much energy is produced (Matteo 1998). This process can be approximated as a magnetic turbulent conductivity $\sigma$ (Bisnovatyi-Kogan 1998, Bisnovatyi-Kogan 1999 Bisnovatyi-Kogan & Lovelace 2002). In (Campbell et al.1998) the authors determine the local dynamo number D for steep function $\alpha$ in the model of the dipole field and they obtain a decision on the function of generation of magneto-rotational instabilities. They found $D_{min}$ for a dynamo in a critical condition when it can only withstand magnetic field diffusion.

The magnetic dissipation in our case has the form $\eta j^2$ and is also represented by the r$\varphi$ - component because the basic movement of electrons is in the direction the $\varphi$-axis and the magnetic viscosity is due to the resistance of radial current $\eta = \alpha_m v_s H$. The rotation can be created and maintained by the radial current even in the absence of kinetic viscosity (Kaburaki 1999). The effects of this resistance, which has appeared as a result of shifting, have to be described by including the vertical stratification of mass density. When the magnetic field increases, the spectrum of unstable modes moves to longer wavelengths. The increase depends strongly on the configuration of the main field- from the toroidal field where the MRI are missing, to the vertical-uniform one where they are most well expressed. Then, by including the vertical scale condition $v_a^2 \succ v_s^2$, there will be no more unsustainable magneto-rotational modes – there will be stability (saturation). In our case asymmetric MR-modes grow most rapidly. Saturation creates a powerful magnetic pressure; it is transported by and along with the instabilities (Parker, PTH, etc.) As a result of that the disk forms a highly magnetized corona. This keeps the balance of energy $E_m \sim E_k$ in the disk and the magnetic field will not be able to suppress rotational effects and put out the MRI.

At this stage the evolution of the disk is mainly determined by the global component of the poloidal field. With a sufficiently low field in a differential rotating system, the linear instabilities can be neglected – with kinetic viscosity, because Relay`s criterion is met; with magnetic viscosity, shifting, which always leads to a dynamo, will not lead to a dynamo action, because when there is a weak vertical global field and radial velocity is different from zero, the dynamo remains in a sub-crytical condition $D \approx D_{cr}$. After the initial increase in magnetic energy it will decrease because the dissipation of the field and the crytical dynamo will be established in a state of self-induction. In this state the dynamo only compensates for losses. (Biskamp 1993, Brandenburg &Subramanian 2004). Then the nonlinear effects in the disk redistribute the energy, the field and the angular momentum.

Coefficient $\alpha_t = \alpha + \alpha_m$: $\alpha_t = \dfrac{v_t^2 + v_{ms}^2}{v_s^2} = \dfrac{\langle \overline{v_r v_\varphi} \rangle}{v_s^2} + \dfrac{\langle \overline{b_r b_\varphi} \rangle}{4\pi\rho v_s^2}$

- The fact that $\alpha$ is a hydrodynamic parameter allows us, for convenience, to set $\alpha=$ constant, as in the standard theory. In this way we reduce the influence of purely hydrodynamic instabilities without ignoring their actions. Under these assumptions, if $\alpha$ is different from the constant, it can only be due to the interactions between the different types of instabilities. On the other hand, it is not necessary to completely ignore the action of HDI.
- The disk rotation produces a stabilizing effect and may suspend the movement of magneto-rotational instabilities, and then the MRI begins to grow. Advection, however, guarantees the movement of instabilities inward and in such a way that it maintains $\alpha_m$ approximately constant.
- $\alpha = const$ ; $\alpha_m = const$ ; (Then the system of equations splits. Entropy is not a parameter of the stream. It characterizes the state of the flow. See part two.)
- In such disks there are conditions for thermal-viscose instability, but it is in a controlled regime, balanced by the magnetic advection. Symbiosis of the mechanisms allows us to reach virial temperature without free fall (Yankova & Filipov 2011) $T \sim T_{vir} = \dfrac{GMm_i}{kr}$.

Finally we should note that we will work in cylindrical *coordinates (r, φ, z; t),* that are more appropriate taking into consideration the flow geometry. In the sense of the assumptions in this model, the thin disk approximation *(H << r)* can be used.

$$\frac{\partial \rho}{\partial t} + \frac{1}{r}\frac{\partial}{\partial r}(r\rho v_r) + \Omega \frac{\partial \rho}{\partial \varphi} = 0 \qquad (8)$$

$$\frac{\partial}{\partial r}(rB_r) + \frac{\partial B_\varphi}{\partial \varphi} = 0 \qquad (9)$$

$$\frac{\partial v_r}{\partial t} + v_r \frac{\partial v_r}{\partial r} + \Omega \frac{\partial v_r}{\partial \varphi} = -\frac{1}{\rho}\frac{\partial p}{\partial r} + \frac{1}{4\pi\rho r}(B_\varphi \frac{\partial B_r}{\partial \varphi} + \frac{rB_z B_r}{H}) \qquad (10)$$

$$\frac{\partial}{\partial t}(\Omega r^2) + v_r \frac{\partial}{\partial r}(\Omega r^2) = \frac{1}{4\pi\rho r}[B_r \frac{\partial}{\partial r}(r^2 B_\varphi) + \frac{r^2 B_z B_\varphi}{H}] + \frac{\vartheta}{r}\frac{\partial}{\partial r}[(r\frac{\partial \Omega}{\partial r})r^2] \qquad (11)$$

$$H = \frac{v_s^2 + \Omega^2 r(r-r_g)}{3B_r B_z}(2\rho r) \qquad (12)$$

$$\frac{\partial B_r}{\partial t} = \frac{1}{r}\frac{\partial}{\partial \varphi}(v_r B_\varphi) - \frac{\partial}{\partial \varphi}(\Omega B_r) + \frac{v_r B_z}{H} + \frac{\eta}{r}\left[\frac{\partial}{\partial r}\frac{\partial B_r}{\partial \varphi} + \frac{\partial}{\partial r}\frac{\partial B_\varphi}{\partial \varphi}\right] +$$
$$+\frac{\eta}{r}\frac{\partial}{\partial r}(r\frac{\partial B_r}{\partial r}) + \frac{\eta}{r^2}\frac{\partial^2 B_r}{\partial \varphi^2} + \eta \frac{2B_r}{H^2} \qquad (13)$$

$$\frac{\partial B_\varphi}{\partial t} = \frac{\partial}{\partial r}(\Omega r B_r) - \frac{\partial}{\partial r}(v_r B_\varphi) + \frac{\Omega r B_z}{H} + \frac{\eta}{r}\left[\frac{\partial}{\partial \varphi}\frac{\partial B_r}{\partial r} + \frac{\partial}{\partial \varphi}\frac{\partial B_\varphi}{\partial r}\right] +$$
$$+\frac{\eta}{r}\frac{\partial}{\partial r}(r\frac{\partial B_\varphi}{\partial r}) + \frac{\eta}{r^2}\frac{\partial^2 B_\varphi}{\partial \varphi^2} + \eta \frac{2B_\varphi}{H^2} \qquad (14)$$

$$\frac{\partial}{\partial r}(rv_r B_z) - \frac{9\eta B_z}{r} = 0 \qquad (15)$$

$$T\frac{\partial S}{\partial t} + \frac{3}{2}v_r \frac{v_a^2}{r} = \vartheta(r\frac{\partial \Omega}{\partial r})^2 - \frac{c}{\chi \rho H^2}(v_s^2 - \frac{v_a^2}{2\alpha_m^2} - RT) + \frac{\eta}{4\pi\rho}(\frac{B_r}{H} + \frac{3\mu}{r^4})^2 \qquad (16)$$

$$p = R\rho T + \frac{B^2}{8\pi} + \frac{aT^4}{3} \qquad (17)$$

**MODIFICATIONS**

We will propose the following modifications to our model built on the basis of the above mentioned assumptions:

- Different instabilities create irregular periodicity in variables of the characteristics of the stream, which should not be ignored and can be used. The disc is not axis-symmetric, but if you divide it into sectors, the following expressions :

$$F_i = F_{i0} \Re_i (x = r/r_0) \exp[\, k_\varphi(x)\varphi + \omega(x)t\,]$$
(18)

could be used for its leading parameters, where $F_{i0}$ are their values of the outer edge of disc $r_0$.

There is a direct dependency of all types of instabilities on the energetic of the disk and respectively feedback in the development of the characteristics of flow, resulting from the action of these instabilities in the flux. The feedback is an expression on the non-obvious dependence on non-linear effects. In (Balbus & Hawley 1998) the feedback for MRI is presented as a term in the equation that describes the development of the dynamo in the linear analysis, but in reality it is not defined.

- By introducing the coefficients $\omega(r)$ and $k_\varphi(r)$, we define feedback in a non-obvious form, for all types of instabilities. If we introduce them in an open view, this would be wrong, because the imposing of any condition would lead to predetermined results.
- The periodic function was not chosen randomly. This choice was based on the analogy with the distribution of Poisson in statistics for random variations of the flow (we do not know beforehand when and how often they occur).

The function view is also connected to the advective nature of the disk [see results in (Filipov et al.2004). and App. 1]. The positive sign of the exponent takes into account the interaction factor. Poisson`s distribution deals with non-interacting particles in an isolated volume. In ideal gas particles do not interact, but this does not apply to vortex particles. Even if we reduce the isolated volume up to the size of the particle the interactions at the boundaries of the vortices must be taken into account. Such interactions stimulate the appearance of new objects in the distribution (Yankova & Filipov 2011).

- The coefficient $\omega(r)$ indicates how often the flow deviates from its course as a result of meeting a structure or a spontaneous disturbance.
- The coefficient $k_\varphi(r)$ is a sine function of the central angle (in radians) between positions of such deflections of one and the same orbit.

We call $\omega(r)$ and $k_\varphi(r)$ – coefficients of meeting. They correlate with wave numbers from the local model. Coefficients do not have specific relations to one given disturbance, because they are global feedbacks. They have a relation to their general distribution in the stream as a whole. Now applying our modifications (App.2) and we get:

$$v_r \frac{\partial \rho}{\partial r} + \rho(\omega + \frac{9\eta}{r^2} + \frac{3v_r}{r} + k_\varphi \Omega) = 0$$
(19)

$$r \frac{\partial B_r}{\partial r} + B_r + k_\varphi B_\varphi = 0$$
(20)

$$\frac{\partial v_s}{\partial r} + \frac{1}{2}(\frac{v_r}{v_s} - \frac{v_s}{v_r})(\omega + \frac{9\eta}{r^2} + \frac{2v_r}{r} + k_\varphi \Omega) - \frac{1}{2}\frac{v_s}{r} = \frac{k_\varphi B_\varphi B_r}{8\pi \rho r v_s} + \frac{B_z B_r}{8\pi \rho H v_s}$$
(21)

$$(\Omega r^2)\omega + v_r \frac{\partial}{\partial r}(\Omega r^2) = \frac{B_r B_\varphi}{2\pi\rho} + \frac{B_r r}{4\pi\rho}\frac{\partial B_\varphi}{\partial r} + \frac{B_z B_\varphi r}{4\pi\rho H} +$$
$$+ \frac{\vartheta}{r}\frac{\partial}{\partial r}\left[\left(r\frac{\partial\Omega}{\partial r}\right)r^2\right] \tag{22}$$

$$H = \frac{v_s^2 + \Omega^2 r(r - r_g)}{3 B_r B_z}(2\rho r) \tag{23}$$

$$B_r(\omega + k_\varphi \Omega) = \frac{2k_\varphi}{r}(v_r B_\varphi) + \frac{v_r B_z}{H} + \frac{\eta}{r}\left[k_\varphi \frac{\partial B_r}{\partial r} + B_r \frac{\partial k_\varphi}{\partial r}\right] -$$
$$- \frac{\eta}{r}\frac{\partial B_r}{\partial r} + \frac{\eta}{r^2}k_\varphi^2 B_r + \eta \frac{2B_r}{H^2} \tag{24}$$

$$B_\varphi\left(\omega + \frac{9\eta}{r^2} + \frac{2v_r}{r} + k_\varphi \Omega\right) = B_r \frac{\partial}{\partial r}(\Omega r) - \Omega B_r + \frac{\Omega r B_z}{H}$$
$$+ \left(\frac{\eta}{r}k_\varphi + \frac{\eta}{r} - v_r\right)\frac{\partial B_\varphi}{\partial r} + \eta\frac{\partial^2 B_\varphi}{\partial r^2} - \frac{\eta}{r^2}k_\varphi B_r + \eta\frac{2B_\varphi}{H^2} \tag{25}$$

$$\frac{\partial v_r}{\partial r} = \frac{9\eta}{r^2} + \frac{2v_r}{r} \tag{26}$$

This is the first set of equations from the split of the system. The modifications that have been introduced, affect the model qualitatively, as they bring essential physical and mathematical benefits:

- The coefficients ω(r) and $k_\varphi$(r), which depend on the distance to the centre, define the feedbacks that are essential to the physics of the object.
- This way of presenting the characteristics allows us to preserve the implicit dependence of the leading parameters on the time and angular coordinate.
- Equations remain nonlinear.
- The modification of the parameters $F_i$, through the coefficients of periodicity depending on the distance to the centre r, permits the reduction of the number of variables. So for purely physical reasons (feedbacks) the PDE system is reduced to an ODE system.

Along the outer radius $r_0$, we turn the ODE system dimensionless (App.3, App. 4):

$$\frac{\partial f_1(x)}{\partial x} + \frac{f_1(x)}{f_2(x)}\left(c_1 f_7(x) + c_2 \frac{f_3(x)f_4(x)}{x^2} + 3\frac{f_2(x)}{x} + c_3 \frac{f_8(x)}{x^{1/2}(x - x_g)}\right) = 0 \tag{27}$$

$$\frac{\partial f_5(x)}{\partial x} + \frac{f_5(x)}{x} + c_4 \frac{f_6(x)f_8(x)}{x} = 0 \tag{28}$$

$$\frac{\partial f_3(x)}{\partial x} + \frac{1}{2}\left(c_5\frac{f_2(x)}{f_3(x)} - \frac{f_3(x)}{f_2(x)}\right)\left(c_1 f_7(x) + c_2 \frac{f_3(x)f_4(x)}{x^2} + 2\frac{f_2(x)}{x} + c_3 \frac{f_8(x)}{x^{1/2}(x-x_g)}\right) -$$

$$-\frac{1}{2}\frac{f_3(x)}{x} = c_6 \frac{f_5(x)f_6(x)f_8(x)}{x f_1(x) f_3(x)} + c_7 \frac{f_5(x)}{x^3 f_1(x) f_3(x) f_4(x)} \tag{29}$$

$$\frac{\partial f_6(x)}{\partial x} = \frac{f_1(x)}{c_8 x f_5(x)}\left(c_1 \frac{f_7(x)x^{\frac{3}{2}}}{x-x_g} + f_2(x)\left[\frac{3}{2}\frac{x^{\frac{1}{2}}}{x-x_g} - \frac{x^{\frac{3}{2}}}{(x-x_g)^2}\right] - 2c_8 \frac{f_5(x)f_6(x)}{f_1(x)} - \right.$$

$$\left. - c_9 \frac{f_6(x)}{x^2 f_1(x) f_4(x)} - c_{10}\frac{\alpha}{x} f_3(x) f_4(x) \frac{\partial}{\partial x}\left[x \frac{\partial}{\partial x}\left(\frac{1}{x^{1/2}(x-x_g)}\right)x^2\right]\right) \tag{30}$$

$$f_4(x) = c_{12}\frac{f_1(x)f_3^2(x)}{f_5(x)}x^4 + c_{14}\frac{f_1(x)}{f_5(x)}\frac{x^4}{(x-x_g)} \tag{31}$$

$$\frac{\partial f_8(x)}{\partial x} = \frac{x}{c_{13}f_3(x)f_4(x)}\left(c_1 f_7(x) + c_3 \frac{f_8(x)}{x^{1/2}(x-x_g)}\right) - \frac{2c_4}{c_{13}}\frac{f_2(x)f_6(x)f_8(x)}{f_3(x)f_4(x)f_5(x)} -$$

$$-\frac{c_{15}}{c_{13}}\frac{f_2(x)}{x^2 f_3(x) f_4^2(x) f_5(x)} + \frac{f_8(x)}{x} + c_4 \frac{f_6(x)f_8^2(x)}{x f_5(x)} - \frac{1}{k_{\varphi 0} x} - \frac{c_4}{k_{\varphi 0}}\frac{f_6(x)f_8(x)}{x f_5(x)} -$$

$$-k_{\varphi 0}\frac{f_8^2(x)}{x} - \frac{c_{11}}{c_{13}}\frac{x}{f_4^2(x)} \tag{32}$$

$$f_6(x)\left(c_1 f_7(x) + c_2 \frac{f_3(x)f_4(x)}{x^2} + 2\frac{f_2(x)}{x} + c_3 \frac{f_8(x)}{x^{\frac{1}{2}}(x-x_g)}\right) =$$

$$= \frac{c_3}{c_4}f_5(x)\frac{\partial}{\partial x}\left(\frac{x^{\frac{1}{2}}}{x-x_g}\right) - \frac{c_3}{c_4}\frac{f_5(x)}{x^{\frac{1}{2}}(x-x_g)} + \frac{c_{17}}{x^{\frac{5}{2}}(x-x_g)f_4(x)} -$$

$$-\frac{c_{16}}{c_4}\frac{f_3(x)f_4(x)f_5(x)f_8(x)}{x^2} + c_{11}\frac{f_3(x)f_6(x)}{f_4(x)} + c_{10}f_3(x)f_4(x)\frac{\partial^2 f_6(x)}{\partial x^2} -$$

$$-\left(f_2(x) - c_{10}\frac{f_3(x)f_4(x)}{x} - c_{13}\frac{f_3(x)f_4(x)f_8(x)}{x}\right)\frac{\partial f_6(x)}{\partial x} \tag{33}$$

$$\frac{\partial f_2(x)}{\partial x} = c_2 \frac{f_3(x)f_4(x)}{x^2} + 2\frac{f_2(x)}{x} \tag{34}$$

## APPROXIMATE SOLUTION

### Iterational method

The boundary problem is inapplicable to the dimensionless system (27-34) we have received, because the problem is not definable. This is due to the fact that it is not possible to guarantee the continuity of the functions of the leading parameters in the defined value area. In this case is more appropriate to set the initial conditions. The initial conditions of the problem are set as boundary conditions of the outer edge of the disc. Then we use iterations (Runge -- Kutta) and as the first step, we apply $\dfrac{\partial f_i(x)}{\partial x} = 0$, $i = 1,\ldots,8$.

$\dfrac{\partial^2 f_6(x)}{\partial x^2} = 0$ because of a leap in $B_\varphi$ of the outer edge, since $B_\varphi$ is generated in the disc. The comfortable initial / boundary conditions $f_i(1)=1$ $f_i''(1)=0$, give us the opportunity to use directly theoretical or observational estimates of the leading parameters for a particular source as initial values $F_{i0}$.

### Non-uniqueness of the decision and the choice (see App. 5)

The second step is a good approximation for the solution of the system, as in this case *(for $f_3(x) = 1$, $f_4(x) = x$)* the rows are rapidly convergent:

$$f_1(x) = \frac{2+c_2}{10x^{10}} + \frac{1+c_1+c_3}{x^{2\frac{1}{2}}}\left[1-(x-x_g)\right] + \frac{8-c_2}{10} \tag{35}$$

$$f_2(x) = (c_2 + 2)(x - 1) + 1 \tag{36}$$

$$f_3(x) = \frac{c_6}{6}x^6 + \frac{c_7 x^5}{5} + \frac{(2+c_2)-c_5(1+c_2)}{2}(x-1) + (1-c_5)\frac{1+c_1+c_3}{2x^{\frac{1}{2}}}(x-x_g-1) +$$
$$+1 - \frac{c_6}{6} - \frac{c_7}{5} \tag{37}$$

$$f_4(x) = (1 - c_{12} - c_{14})x \tag{38}$$

$$f_5(x) = \frac{1+c_4}{2x^2} + \frac{1-c_4}{2} \tag{39}$$

$$f_6(x) = \frac{2+3\alpha c_{10}}{4c_8 x^{17/2}}(x-x_g-1) - \left[\frac{1+c_1}{c_8 x^8} + \frac{2\alpha c_{10}-1}{c_8 x^{15/2}}\right]\frac{1}{(x-x_g)} +$$

$$+ \frac{\alpha c_{10}}{c_8 x^{13/2}}\frac{1}{(x-x_g)^2} + \frac{c_9}{3c_8 x^3} + \frac{1}{x^2} + \frac{c_1+\alpha c_{10}}{c_8} - \frac{c_9}{3c_8}$$

(40)

$$f_7(x) = \left[\frac{4c_{10}-c_2+c_{11}-2c_{13}}{c_1} - \frac{c_{16}}{c_1 c_4}\right]\frac{1}{x} - \left[\frac{c_3}{c_1}\frac{1}{x^{1/2}} + \frac{c_3}{2c_1 c_4 x^{1/2}} - \frac{c_{17}}{c_1 x^{3/2}}\right]\frac{1}{x-x_g} -$$

$$- \frac{c_3}{c_4 c_1}\frac{x^{1/2}}{(x-x_g)^2} + 1 + \frac{c_2}{c_1} + \frac{c_3}{c_1} - \frac{4c_{10}}{c_1} - \frac{c_{11}}{c_1} + \frac{2c_{13}}{c_1} - \frac{c_{17}}{c_1} + \frac{3c_3+2c_{16}}{2c_4 c_1}$$

(41)

$$f_8(x) = \left(c_4+1-k_{\varphi 0} - \frac{c_4+1}{k_{\varphi 0}} - \frac{1+2c_4+c_{11}}{c_{13}}\right)(x-1) + \frac{c_{15}}{c_{13}x} - \frac{1+c_1+c_3}{c_{13}x^{1/2}}(x-x_g-1) +$$

$$+ \frac{c_{13}-c_{15}}{c_{13}}$$

(42)

Here $f_1(x)$, $f_2(x)$, $f_3(x)$, $f_4(x)$, $f_5(x)$, $f_6(x)$, $f_7(x)$, $f_8(x)$ are respectively the dimensionless functions of the parameters $\rho$, $v_r$, $v_s$, $H$, $B_r$, $B_\varphi$ of the flow, and the coefficients of meeting $\omega$, $k_\varphi$. And $c_i$ are dimensionless combinations of parameter values at the outer edge of the disc (App. 4).

**FOLLOWING THE EVOLUTION IN THE MOMENTS WE HAVE SELECTED: t=1P~$\Omega_0^{-1}$ AND t ≈ 0**

The model gives solutions in different evolutionary moments, but in this article the results are treated only for the two crucial points t=1P~$\Omega_0^{-1}$ and t ≈ 0. The graphical images of the leading parameters help us interpret the results and register the appearance of spirals, microstructures and short-living rings in the inner region as well as to observe the initial spreading of the disk.

Below one can see dimensionless distributions $f_i(x, \varphi)$ of the main physical parameters (characteristics) in the plane of the disk (X, Y), where $X = x.\cos\varphi$ and $Y = x.\sin\varphi$. They describe the decision to the 2D structure of the disc at the moment t=1P after the spreading.

The solution for the moment t ≈ 0 is obtained and has the form $f_i(x)exp(-\omega_0 f_7(x)/\Omega_0)$ (Iankova & Filipov 2005, Iankova 2009), (* this is due to the fact that around this moment there are no instabilities in the disc yet (Hawley & Balbus 2002)). The coefficient $\omega(r)$ is not defined and $k_\varphi(r)$ is zero.

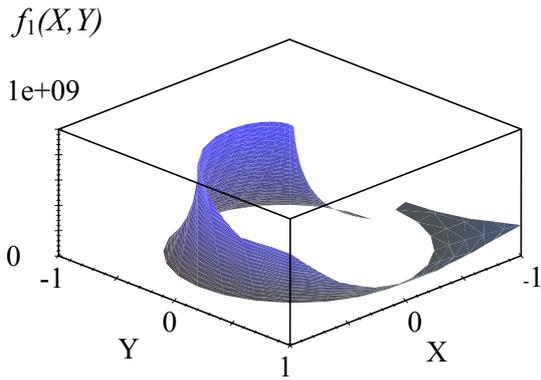 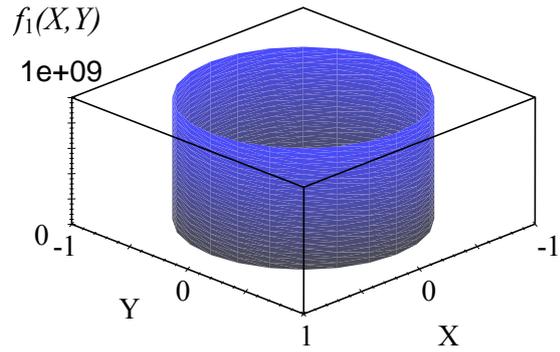

**Fig.1a:** Dimensionless function of the density distribution $f_1(X, Y)$ in the non-axe-symmetric MHD model. The spiral could be seen as a leap of density function.

**Fig.1b:** That is a dimensionless function of density distribution $f_1(X, Y)$ in the disc at $t \approx 0$.

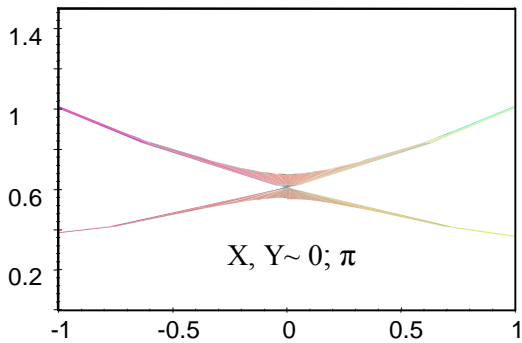 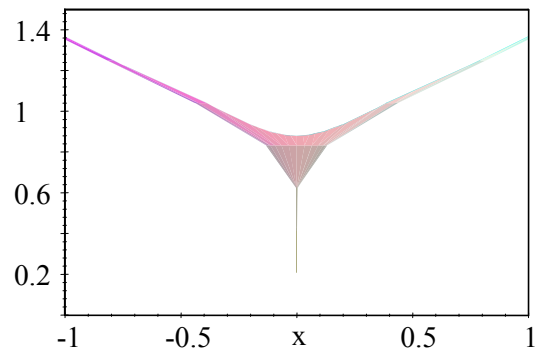

**Fig.2a:** Profile of the dimensionless function of the radial velocity $f_2(X, Y)$ at the fixed angle coordinate, at $t=1P\sim\Omega_0^{-1}$.

**Fig.2b:** Profile of the dimensionless function of the radial velocity $f_2(X, Y)$ at the fixed angle coordinates, at $t \approx 0$.

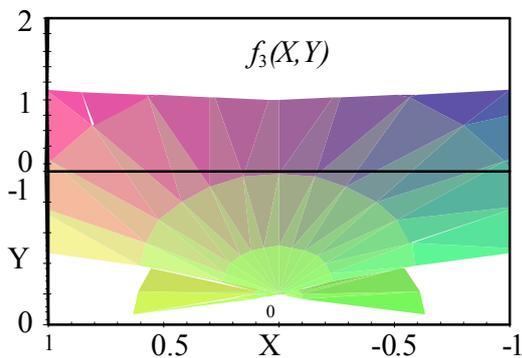 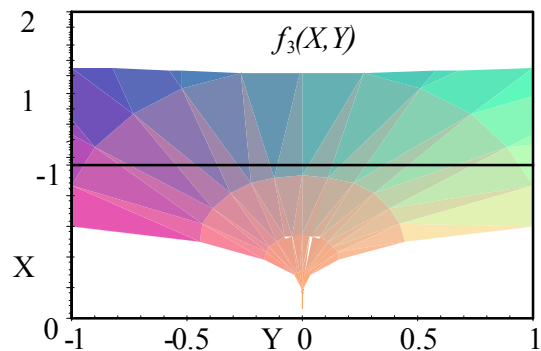

**Fig.3a:** Dimensionless function of the sound speed distribution $f_3(X, Y)$ in the disc at $t \approx \Omega_0^{-1}$.

**Fig.3b:** Initial distribution of the dimensionless sound speed $f_3(X, Y)$ in the disc plane. The function smoothly decreases. At the initial spreading the branch of the seals is missing.

Fig.1a is a surface distribution of the equatorial density. The figure shows the presence of a spiral. The split in the surface of the function marks beginning $x \approx 0.8 \div 0.9$ of the spiral. The smooth increase of the equatorial density in layer $0.9 \div 1.0x$, shows that the initial front in the disk, formed by the inflow, is separated from the spiral (similar behaviour can be observed in the numerical results of Bisikalo (Bisikalo et al.2001; a, b). In the following simulations, the group shows that at temperatures higher than $10^6$ the hot disks have only one tidal spiral. The other spiral is destroyed under the impact of the hot line.) At moment $t \approx 0$ (Fig.1b), matter is accumulated mainly into the ring at orbit $x \sim 0.8$ and is slowly spreading inside.

Fig.2 shows two branches of the function of the radial velocity. The profile indicates that the two branches are individual along the whole disk (2a). Radial velocity increases at one of the branches while at the same time it decreases at the other. This means that the fluid moves freely and independently in both directions. Such behaviour clearly indicates the presence of micro-vortices in the disk (Some of them are structures! If there was only turbulence, then speed would not have had two defined priority directions. $V_r$ would have had only one direction, definable by the summary chaotic effect).

The function of the radial velocity (2b), does not have only one meaning for $x<0.3$. This implies that the flow of the plasma is in the shape of weak unstable accretion threads. This means that the disc shows its asymmetric nature even at this stage of development.

The form of sound speed (fig.3a) indicates that one period after the spreading of the disk in the inner regions for $x<0.6$ the formation of higher density rings (compared to the density of the environment) begins. Continuity of the surfaces of the two branches shows that the rings are mobile formations flowing into each other (if the function was interrupted anywhere, this is would have been impossible). These rings crawl into the inner zones of the disk. They are not cause by self-gravity and therefore do not contradict the initial assumption of its absence.

The sound speed at moment $t \approx 0$ (fig.3b) decreases slowly. Plasma still flows smoothly into the inner regions.

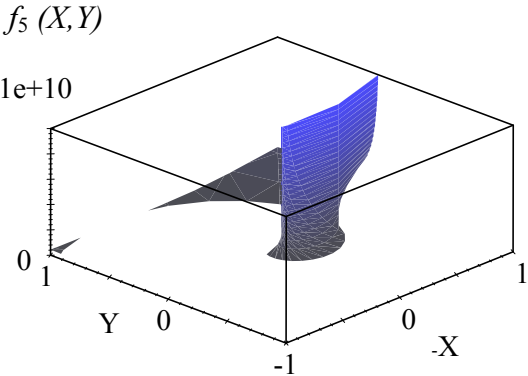
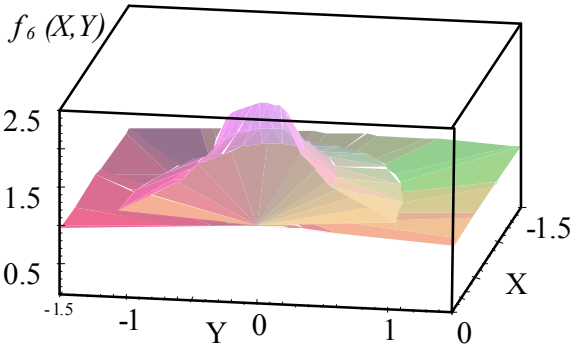

**Fig.4a:** That is a dimensionless function of radial component of the magnetic field distribution $f_5(X, Y)$.

**Fig.4b:** That is a dimensionless function of azimuth component of the magnetic field distribution $f_6(X, Y)$.

The split in distribution of the radial field (Figure 4a), give us possibility to evaluate where and for what increase in magnetic field, MRI can be realized. A jump in the function was observed at $x \sim 0.5$. Therefore, MHD instability is expected below this radius. Smooth increasing of the radial field above this radius corresponds to the weak development of the azimuth component (Fig. 4b) and the field as a whole.

Slow increase in $B_\varphi$ is very important in the case. Fig. 4b shows that for each point of the 2D disk, the magnitude of the field has at least three possible values, and at least two of them are low harmonics near $B_{\varphi 0}$. This means that the azimuthal field energy is diffused additionally. These are factors to retain dynamo in the sub-critical regime and allowing nonlinear effects to occur.

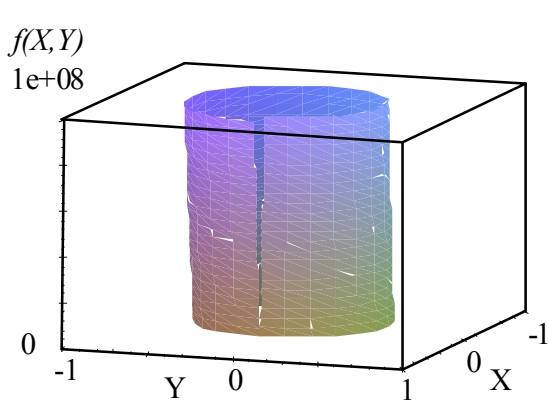 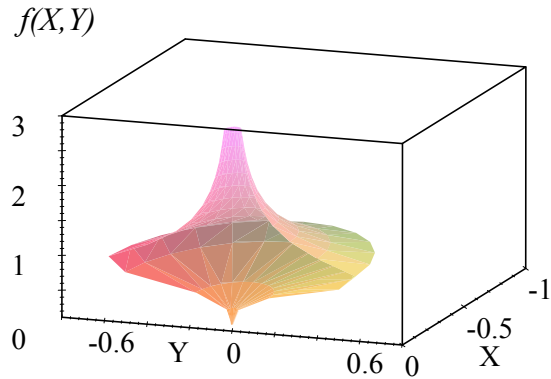

**Fig.5a:** That is a dimensionless function of magneto-sonic velocity $f(X, Y)$.

**Fig.5b:** Initial distribution of the dimensionless magneto-sonic velocity $f(X, Y)$ in the disc plane.

The magnetic sound speed (fig.5a) with small increases resembles the development of $B_\varphi$. Since, however, the magnetic sound speed also contains the contribution of the other components; it has a much faster increase (compared to $B_\varphi$). Around moment $t \approx 0$ it was very poorly developed.

**CONCLUDING COMMENTS**

In this paper we present a developed model of magneto-hydrodynamics of an accretion disk based on a specific advective hypothesis. The aim of the paper is to describe the accretion flow in detail, in its capacity of an open astrophysical system and to interpret the mechanisms of disk self-structuring in an accurate way.

The model allows us to: observe the evolution of the disc; investigate the emergence of instability in it; study the generation of its corona.

In the modelling process we have introduced "coefficients of meeting". They give expression of the feedbacks, resulting from the impact of nonlinear effects on the nature of the flow. That gives the model both physical and mathematical advantages: equations remain nonlinear; the number of variables is reduced without the loss of physical dependencies.

We have also obtained a solution for the radial structure of the disc. This solution has been investigated for two crucial moments of disk evolution. When the results for the disk self-organization were analyzed, we found the appearance of a spiral, the presence of micro-vortices (microstructures) and the formation of pseudo-rings in the inner regions.

The heat balance and local conditions in registered rings with higher density also have an important role to play in the auto-structuring of the disk. This role, however, will be the subject of further studies.

The model provides a wide scope of applications with real sources and it can be used in the future to investigate problems such as: stability, formation of the disk corona, advection in the disc and interaction in the corona-disk system in objects that have not been studied sufficiently.

*Acknowledgements*. We appreciate the support of our colleagues: Assoc. Prof. Sv. Zhekov and PhD S. Asenovsky from SRTI-BAS, for the technical support and layout of this paper.

%%%%%%%%%%%%%%%%%%%%%%%%%%%%%%%%%%%%%%%%%%%%%%%%%%%

**Appendix 1**

Exponent:

The results presented in (Filipov et al.2004) show self-similar solutions for the structure of the disk with and without advection (but in both cases without a MF). Irrespective of the similar approach to both discs, they showed different behavior. The distributions in the advective disk develop in an exponential time scale, while the other disk shows degrees characteristics. This feature of the results is not directly related to the aim of article (Filipov et al.2004). It, however, gives us an adequate idea of the specific modification in the model presented in this article. The advective nature of the disc implies the exponential character of the development. So as to avoid the doubt that there is a contradiction in the model, we should point out that the intensity of development of advection depends on the argument of the function and not on its type. Yet again we emphasize the fact that advection is not the dominant mechanism here.

**Appendix 2**

The modification "coefficients of meeting" is applied to the system (8-17) resulting in the system (19-26). This is shown through the second equations in the systems:

$$F_i = F_{i0} \Re_i (x = r/r_0) \exp[k_\varphi(x)\varphi + \omega(x)t],$$

$$B_r(x)\exp[k_\varphi(x)\varphi + \omega(x)t] \qquad B_\varphi(x)\exp[k_\varphi(x)\varphi + \omega(x)t]$$

$$\frac{\partial}{\partial r}(rB_r) + \frac{\partial B_\varphi}{\partial \varphi} = 0 \quad (9) \qquad r\frac{\partial B_r}{\partial r} + B_r + k_\varphi B_\varphi = 0 \quad (20)$$

$$r\frac{\partial}{\partial r}B_r(x)\exp[k_\varphi(x)\varphi + \omega(x)t] + B_r(x)\exp[k_\varphi(x)\varphi + \omega(x)t]\frac{\partial r}{\partial r} +$$
$$+ k_\varphi(x)B_\varphi(x)\exp[k_\varphi(x)\varphi + \omega(x)t] = 0$$

**Appendix 3**

According to the essence of the model, the characteristics of the flow, changes in the same way, independently from one another when the flow meets a structure or disturbance.

$$F_i = F_{i0} \Re_i (x = r/r_0) \exp[k_\varphi(x)\varphi + \omega(x)t] =$$
$$= F_{i0} f_i [x, k_\varphi(x), \omega(x)] = F_{i0} f_i(x)$$

The relations for t and φ accepted in this way, have coefficients of periodicity depending on the distance to the center r (the feedbacks are closed- locked). Along the coordinates r and z the functions retain their non-periodic nature. The modification in parameters $F_i$, reduces the number of variables. System (19-26) is a dimensionless one along the outer edge $r_0$ resulting in system (27-34). This is shown through the second equations in the systems:

$$\frac{F_i}{F_{i0}} = f_i(x), \quad (20) \qquad r\frac{\partial B_r}{\partial r} + B_r + k_\varphi B_\varphi = 0$$

$$r_0 x \frac{\partial B_{r0} f_5(x)}{\partial r_0 x} + B_{r0} f_5(x) + k_{\varphi 0} B_{\varphi 0} f_6(x) f_8(x) = 0$$

$$\frac{\partial f_5(x)}{\partial x} + \frac{f_5(x)}{x} + c_4 \frac{f_6(x)f_8(x)}{x} = 0 \qquad (28).$$

**Appendix 4**

| $c_1 = \dfrac{r_0 \omega_0}{v_{r0}}$ | $c_2 = \dfrac{9 v_{s0} H_0}{v_{r0} r_0}$ | $c_3 = \dfrac{\Omega_0 r_0 k_{\varphi 0}}{v_{r0}}$ | $c_4 = \dfrac{k_{\varphi 0} B_{\varphi 0}}{B_{r0}}$ |
|---|---|---|---|
| $c_5 = \dfrac{v_{r0}^2}{v_{s0}^2}$ | $c_6 = \dfrac{k_{\varphi 0} B_{\varphi 0} B_{r0}}{8\pi \rho_0 v_{s0}^2}$ | $c_7 = \dfrac{B_{z0} B_{r0} r_0}{8\pi \rho_0 H_0 v_{s0}^2}$ | $c_8 = \dfrac{B_{r0} B_{\varphi 0}}{4\pi \rho_0 r_0 \Omega_0 v_{r0}}$ |
| $c_9 = \dfrac{B_{z0} B_{\varphi 0}}{4\pi \rho_0 H_0 \Omega_0 v_{r0}}$ | $c_{10} = \dfrac{v_{s0} H_0}{v_{r0} r_0}$ | $c_{11} = \dfrac{2 v_{s0} r_0}{v_{r0} H_0}$ | $c_{12} = \dfrac{2 \rho_0 v_{s0}^2}{3 B_{z0} B_{r0}}$ |
| $c_{13} = c_{10} k_{\varphi 0}$ <br><br> $c_{16} = c_{10} k_{\varphi 0}^2$ | $c_{14} = \dfrac{2 \rho_0 r_0^2 \Omega_0^2}{3 B_{z0} B_{r0}}$ | $c_{15} = \dfrac{r_0 B_{z0}}{H_0 B_{r0}}$ | $c_{17} = \dfrac{\Omega_0 r_0^2 B_{z0}}{v_{r0} H_0 B_{\varphi 0}}$ |

**Appendix 5**

Options:

1. $f_3(x) = const$ ; $f_4(x) = x$

2. $f_3(x) = x^{-1}$ ; $f_4(x) = x^2$

3. $f_3(x) = x^{-2}$ ; $f_4(x) = x^3$.

When we use relations $F_i$ for the leading parameters of the disc, we obtain intermediate results (Iankova & Filipov 2005). After the first step we have freedom to choose between three possible solutions. We selected the most suitable one for our studies (1.) - it fits best our need for rapid growth of the radial field and the weak alteration of the azimuthal field; it also corresponds to the choice of a gravity center- a black hole. On the basis of (Quataert &Narayan 1998) we know that $B_r$ intensifies magneto-hydrodynamic turbulence and $B_\varphi$ suppresses it.

This is how we obtain the global solutions for the structure and evolution of the disk in this case. The other two approximations correspond to the following cases: the distension in torus or the destruction of the disk in the propeller system. They are not treated further.